\documentclass{article}
\begin{document}                  

\title{$A$-model and generalized Chern-Simons theory}
\author{A. Schwarz\thanks{Partly supported by NSF
grant No. DMS 0204927.} \\
{\it  Department of Mathematics,}\\
{\it University of California,}\\
{\it Davis, CA 95616, USA}\\
schwarz@math.ucdavis.edu}
\date{}
\maketitle

\begin{abstract}                     

 The relation between open topological strings and  
Chern-Simons theory was discovered 
by E. Witten. He proved that $A$-model on $T^*M$ where 
$M$ is a three-dimensional manifold 
is equivalent to  Chern-Simons theory on $M$ and that 
$A$-model on arbitrary Calabi-Yau 3-fold 
is related to  Chern-Simons theory with instanton 
corrections. In present paper we discuss
multidimensional generalization of these results. 
\end{abstract}

                                     0. {\bf 
Introduction}

In present paper we analyze the relation between 
multidimensional $A$-model of open topological strings 
and 
generalized  Chern-Simons theory. Such a relation was 
discovered by E. Witten \cite {[16]} in 
three-dimensional 
case; we generalize his results. Our approach is based 
on   rigorous mathematical  results of \cite {[3]}, 
\cite {[4]},
\cite {[5]}, \cite {[8]}, \cite {[11]}; 
in 
 three-dimensional case it gives mathematical 
justification of some of Witten's statements. 

 In modern language Witten considers $A$-model in 
presence of a stack of $N$ coinciding $D$-branes 
wrapping 
a Lagrangian submanifold $M$. In the neighborhood of  
Lagrangian submanifold a symplectic manifold $V$ 
looks like $T^*M$. In the case $V=T^*M$, dim$M=3$ Witten 
shows that $A$-model is equivalent to  Chern-
Simons theory on $M$. He considers also the case when 
$V$ is a Calabi-Yau 3-fold and shows that in this 
case Chern-Simons action functional on $M$ acquires 
instanton corrections. 

We remark that one can analyze instanton corrections to 
Chern-Simons functional combining results by 
Fukaya \cite {[5]} and Cattaneo-Froehlich-Pedrini \cite 
{[3]} and that 
this approach works also in multidimensional case. 

  To study the origin of  Chern-Simons functional and 
its generalizations one can replace the stack of $N$ 
coinciding $D$-branes by $N$  Lagrangian submanifolds 
depending on $\varepsilon$ and tending to the same 
limit as $\varepsilon \rightarrow 0$. This situation was 
studied by Fukaya-Oh \cite {[8]} and 
Kontsevich-Soibelman \cite {[11]}; 
we will show that the appearance of  Chern-Simons 
functional follows from their results. 

                                          1.  {\bf 
Generalized  Chern-Simons theory}

  Multidimensional generalization of  Chern-Simons 
theory can be constructed in the following way. 
We consider differential forms on $d$-dimensional 
compact manifold $M$ taking values in Lie algebra 
${\cal G}$. One assumes that ${\cal G}$ is   equipped 
with invariant inner product. We will restrict 
 ourselves to the only case we need: ${\cal G}=gl(N)$; 
then  invariant inner product can be defined 
as $<a,b>={\rm Tr}ab$ where ${\rm Tr}$ denotes the trace 
in vector representation of ${\cal G}=gl(n)$. 
The graded vector space $\Omega ^* (M)\otimes {\cal G}$ 
of such forms will be denoted by ${\cal E}$. 
The bilinear form $<C,C^{\prime} > = \int_M {\rm Tr} 
C\cdot C^{\prime}$specifies an odd symplectic structure 
on ${\cal E}$ 
if ${\rm dim}M$ is odd and even symplectic structure if 
${\rm dim}M$ is even. 

The generalized  Chern-Simons functional $CS(C)$ is 
defined by the standard formula
\begin {equation} 
{\cal S}(C)=CS(C)={1\over 2} \int _M  {\rm 
Tr}CdC+{1\over 3} \int _M {\rm Tr}C [C,C]
\end {equation}
where $C\in {\cal E}=\Omega ^*(M) \otimes gl (n)$ and 
$d$ stands for the de Rham 
differential . We can replace $d$ in (1) by the 
differential $d_A$ corresponding to flat 
connection $A$; corresponding functional will be denoted 
by ${\cal S}_A$. Notice 
that the functional ${\cal S}_A$ for arbitrary flat 
connection in trivial vector bundle 
can be obtained from the functional (1) with  the 
standard  de Rham differential by 
means of shift of variables. It is easy to see that for 
any solution $A$ of 
equation $dA +{1\over 2} [A,A]=0$ we  have 
\begin {equation} 
{\cal S}(C+A)={\cal S}(A)+{1\over 2} \int _M  {\rm 
Tr}(CdC+C[A,C])+{1\over 3} \int _M {\rm Tr}C [C,C]
\end {equation}
If $A$ is a $1$-form such a solution determines a flat 
connection and (2) coincides (up to a constant ) with 
the 
corresponding action functional. This remark permits us
to reduce the study of Chern-Simons functional with flat
connection to the study of functional (1).

In the case  when ${\rm dim}M$ is odd ${\cal E}$  is an 
odd symplectic space hence 
we can define an odd  Poisson bracket on the space of 
functionals on ${\cal E}$  (on the space 
of preobservables); the functional ${\cal S}_A$ obeys 
the BV classical master equation 
$\{ {\cal S}_A,{\cal S}_A \}=0$ and therefore can be 
considered as an action 
functional of classical mechanical system in 
BV-formalism. Corresponding equations of 
motion have the form $$d_A C+{1\over 2} [C,C] =0.$$

The functional ${\cal S}$ determines an odd differential 
$\delta$ on the algebra of preobservables by 
the formula $\delta ({\cal O})=\{ {\cal S}, {\cal O}\}$; 
homology of $\delta$ are identified with classical 
observables.

In the case of even-dimensional manifold $M$ the 
functional (1) has an interpretation in 
terms of BFV-formalism. The Poisson bracket on the space 
of functionals on ${\cal E}$  (on the space 
of preobservables) is even; the operator $\delta$ can be 
interpreted as BRST operator and its homology 
as classical observables. 

The generalized  Chern-Simons action functional (1) was 
considered in \cite {AKSZ},\cite {[15]}  in the 
framework of BV 
sigma-model. In the definition of BV sigma-model we  
consider the space ${\cal E}$ of maps of 
${\Pi}TM$, where $M$ is a $d$-dimensional manifold into 
(odd or even) symplectic $Q$-manifold $X$. 
(One says that a supermanifold equipped with an odd 
vector field obeying $\{ Q,Q\}=0$ is a 
 $Q$-manifold. De Rham differential specifies the 
structure of  $Q$-manifold on ${\Pi}TM$.) The space 
of maps of  $Q$-manifold into a  $Q$-manifold also can 
be regarded as a  $Q$-manifold. From  the 
other side using the volume element on ${\Pi}TM$ and 
symplectic structure on $X$ we can define odd 
or even  symplectic structure on ${\cal E}$. These facts 
permit us to consider BV or BFV theory 
where fields are identified with functionals on ${\cal 
E}$.

Numerous topological theories can be obtained as 
particular 
cases of BV sigma-model. It was shown in \cite {AKSZ} 
that $A$-model and 
$B$-model can be constructed this way.

To obtain generalized Chern-Simons 
theory from BV-sigma model we should take $X=\Pi {\cal 
G}$ in this 
construction. (If $\cal G$ is a Lie algebra we can 
consider 
$\Pi {\cal G} $ as a $Q$-manifold where $Q$ is a vector 
field 
${1\over 2}f^{\gamma}_{\alpha 
\beta}c^{\alpha}c^{\beta}{\partial \over \partial 
c^{\gamma}}$. We use 
the notation $c^{\alpha}$ for coordinates in $\Pi {\cal 
G}$ corresponding to the basis $e_{\alpha}$ in $\cal G$; 
structure constants of $ \cal G$ corresponding to this 
basis are denoted by $f^{\alpha}_{\beta \gamma}$. 
An invariant inner product on $\cal G$ specifies a 
symplectic structure on $\Pi {\cal G}$.)

In \cite {[K]} Kontsevich constructed a multidimensional 
generalization 
of perturbation series for standard Chern-Simons.
It was shown in \cite {[15]} that the perturbation 
theory for generalized
Chern-Simons theory coincides with Kontsevich 
generalization.
It is important to emphasize that usual correlation 
functions of
multidimensional Chern-Simons theory are trivial, 
however, one can define non-trivial cohomology classes 
of some space that play the role of generalized 
correlation functions. (In \cite{[K]} this space was 
related to the classifying space of diffeomorphism group 
of $M$, in \cite {[15]} it was interpreted as moduli 
space of gauge conditions in the corresponding BV 
sigma-model.)

Notice that one can construct Chern-Simons functional 
for every differential associative ${\bf Z}_2$-graded 
algebra ${\cal A}$ equipped with invariant inner product 
$< , >$. (We assume that the algebra is unital; 
then the invariant inner product can be written in terms 
of trace: $<a,b>={\rm tr} ab$.) For every $N$ we 
define the associative algebra ${\cal A}_N$ as tensor 
product ${\cal A} \otimes {\rm Mat}_N$ where 
${\rm Mat}_N$ stands for the matrix algebra. We define 
Chern-Simons functional for $A\in {\cal A}_N$ 
by the formula
$$ CS(A)={1\over 2} {\rm tr}AdA+{2\over 3}  {\rm 
tr}A^3={1\over 2} {\rm tr}AdA+{1\over 3} {\rm 
tr}A[A,A]$$
(Notice that we need really only the super Lie algebra 
structure defined by the super commutator in 
the associative algebra ${\cal A}_N$.)

The functional $CS$ coincides with (1) in the case when 
${\cal A}$ is the algebra $\Omega (M)$ of differential 
forms on manifold $M$ equipped with a trace ${\rm 
tr}C=\int_M C$.

The construction of $CS$  functional can be generalized 
to the case when ${\cal A}$ is an 
$A_{\infty}$-algebra equipped with invariant inner 
product. Recall that the structure of $ 
A_{\infty}$-algebra 
 ${\cal A}$ on a ${\bf Z}_2$-graded space is specified 
by means of a sequence $^{(k)}m$ of operations; in a 
coordinate system the operation  $^{(k)}m$ is specified 
by a tensor $^{(k)}m^a_{a_1,...,a_k}$ having one upper 
index and $k$ lower indices. Having an  inner product we 
can lower the upper index; invariance of  inner product 
means that the tensor  $^{(k)}\mu _{a_0,a_1,...,a_k} 
=g_{a_0 a} m^a_{a_1,...,a_k}$ is cyclically symmetric 
(in graded sense). The Chern-Simons functional  can be 
defined on ${\cal A}\otimes {\rm Mat }_N$ 
by means of tensors  $^{(k)} \mu $; see \cite {[12]} for 
details.  
Notice that two quasiisomorphic $A_{\infty}$-algebras 
are
physically equivalent (i.e. corresponding Chern-Simons
functionals lead to the same physical results).

A differential associative algebra can be considered as 
an $A_{\infty}$-algebra where only 
operatious $^{(1)} m$ and $^{(2)} m$ do not vanish; in 
this case both definitions of Chern-Simons functional 
coincide.

        2. {\bf Observables of  Chern-Simons theory.}

If  Chern-Simons theory is constructed by means of 
associative graded differential algebra ${\cal A}$ 
with  inner product  it is easy to check that classical 
observables of this theory correspond to cyclic
cohomology of ${\cal A}$. This fact is equivalent to the 
statement that infinitesimal deformations 
of ${\cal A}$ into $A_{\infty}$-algebra with  inner 
product are labelled by cyclic cohomology $HC({\cal A})$ 
of ${\cal A}$ \cite {[17]}. (Recall, that classical 
observables 
are related to infinitesimal deformations of the 
theory.) Algebra ${\cal A}$ determines Chern-Simons 
theory  for all $N$;the 
observables we were talking about were defined  for 
every $N$.

As we mentioned the generalized Chern-Simons theory 
corresponds to the algebra of differential 
forms $\Omega ^*(M)$ with de Rham differential. It is 
well known \cite {[2]}, \cite {[10]} that cyclic  
cohomology of 
this algebra are related to equivariant homology of loop 
space $L(M)$. More precisely, there 
exists a map of equivariant  homology $H_{S^1}(L(M))$ 
into  cyclic cohomology 
 $HC(\Omega ^* (M),d)$; if $M$ is simply connected this 
map is an isomorphism. 

Recall that the loop space $LM$ is defined as a space of 
all continuous maps of the circle 
$S^1={\bf R}/{\bf Z}$ into $M$; the group $S^1$ acts on 
$LM$ in obvious way: 
$\gamma (t)\rightarrow \gamma (t+s)$. It will be 
convenient to modify the definition of 
$LM$ considering only piecewise differential maps; this 
modification does not change 
the homology. 

Instead of equivariant homology of 
$LM$ one can consider homology of the space of closed 
curves (string space) $SM$ obtained 
from $LM$ by means of factorization with respect to 
$S^1$. The manifold $M$ is embedded in 
$LM$ and in $SM=LM/S^1$ as the space of constant loops; 
excluding constant loops from 
consideration we can identify $S^{1}$-equivariant 
homology of $LM\setminus M$ with 
 homology of $SM\setminus M$. (In general $S^1$- 
equivariant homology of over real 
numbers can be identified with  homology of quotient 
space if all stabilizers are finite.)
{\footnote { The space $SM\setminus M$ is an 
infinite-dimensional orbifold with  orbifold 
points corresponding to $n$-fold curves. See \cite 
{[14]} for the 
first analysis of  orbifold  structure of 
$SM$ and calculation of homology of $SM$ in simple 
cases.}}

Following \cite {[4]} we will use the term "string  
homology" and 
the notation ${\cal H}_*M$ for the homology of string 
space $SM$.

The homomorphism of  ${\cal H}_\ast M$ into the space of 
observables of  
Chern-Simons theory can be described in the following 
way \cite {[3]}.

Let us consider the standard symplex 
$\Delta _n=\{(t_1,...,t_n)\in {\bf R}^n|0\leq 
t_1\leq...\leq t_n\leq 1\}$ and evaluation maps 
$ev_{n,k}:\Delta_n\times LM\rightarrow M$ that transform 
a point 
$(t_1,...,t_n,\gamma)\in \Delta _n\times LM$ in 
$\gamma(t_k)$ (here $1\leq k\leq n)$. 
Using these maps we can construct a differential form on 
$LM$ by the formula
$${ h}(C)={\rm Tr} 
\int_{\Delta_n}ev^*_{n,1}C...ev^*_{n,n}C$$
where $C\in {\cal E}=\Omega ^*(M)\otimes {\rm Mat}_N$ is 
a differential form on $M$ 
taking values in $N\times N$ matrices. We obtain a map 
of the space of fields of 
 Chern-Simons theory into $\Omega ^*(LM)$.

 The form $h(C)$ descends to the string space $SM$. If 
$a$ is 
a singular chain in $SM$, then 
\begin {equation} 
\rho _a(C)=\int _ah(C)
\end {equation}
specifies a functional on the space ${\cal E}$ of fields 
(a preobservable of  
Chern-Simons theory ). It follows from results of \cite 
{[3]} 
that 
\begin {equation}
\delta \rho _a=(-1)^n \rho_{\partial a},
\end {equation}
where $\partial a$ stands for the boundary of the chain 
$a$.
This means, in particular, that in the case when $a$ is 
a cycle in the homology
of $SM$ (in the string homology) $\rho_a$ is an  
observable and that two 
homologous cycles specify equivalent  observables. We 
obtain a map of string homology ${\cal H}_*M$  
into the space of observables of  Chern-Simons theory on 
$M$.

                  {\bf 3.String bracket.}

Let us describe some operations in homology of loop 
space $LM$ and string space $SM$ 
that were introduced in \cite {[4]}. 

The most fundamental of these operation is the loop 
product on the loop space.  It assigns 
(under some transversality assumptions) an 
$(i+j-d)$-dimensional chain $a\bullet b$ in $LM$ 
to $i$-dimensional chain $a$ and $j$-dimensional chain 
$b$. To construct $a\bullet b$ 
one first intersects in $M$ the chain of marked points 
of $a$ with  the chain of marked 
points of $b$ to obtain an $(i+j-d)$-dimensional chain 
in $M$ along which the  marked 
points of $a$ coincides with  the marked points of $b$. 
Now one defines the chain 
$a\bullet b$ by means of concatenation of the loops of $a$ 
and the loops of $b$ having 
common  marked points.

The operator $\Delta $ on the chains of the loop space 
$LM$ transforms an  
 $i$-dimensional chain $a$ into $(i+1)$-dimensional 
chain $\Delta a$ obtained by 
means of circle action on $LM$. 

The bracket $\{a,b \}$ of $i$-dimensional chain $a$ in 
$LM$ and  $j$-dimensional chain 
$b$ in $LM$ is an $(i+j+1)$-dimensional chain that can 
be defined by the formula
\begin{equation}
\{a,b \}=(-1)^i\Delta (a\bullet b)-(-1)^i\Delta a\bullet 
b-a\bullet \Delta b
\end {equation}
All these operations descend to homology of $LM$; the 
homology becomes a 
Batalin-Vilkovisky algebra \cite {[9]}, \cite {[13]} 
with 
respect to them.

A natural map of $LM$ onto $SM$ (erasing the marked 
point) determines a homomorphism 
proj of chain complexes. An $i$-dimensional chain in 
$SM$ can be lifted to 
$(i+1)$-dimensional chain in $LM$ (we insert marked 
points in all possible 
ways); corresponding homomorphism of chain complexes 
will be denoted by lift.

The string bracket of two chains in $SM$ can be defined 
by the formula
\begin{equation}
[a,b]={\rm proj}({\rm lift} b\bullet {\rm lift}a).
\end {equation}
 If ${\rm dim } a=i,\ \  {\rm dim}b=j$, then ${\rm 
dim}[a,b]=i+j-d+2$
This bracket descends to homology of $SM$ (to string 
homology),defining a graded 
Lie algebra. 
The above definition of bracket agrees with \cite {[3]}; 
in the 
definition of \cite {[4]} $a$ and $b$ are interchanged.

As we know, there exists a map of string homology into  
the space of observables. The 
main result of \cite {[3]} is a theorem that this map is 
compatible with Lie algebra 
structures on string homology and on  the space of 
observables:
\begin {equation}
\{ \rho_a,\rho_b\}=\rho_{[a,b]},
\end {equation}
where $\{ , \}$ stands for the Poisson bracket. 

It is important to notice that (7) remains correct if 
$a$ 
and $b$ are arbitrary chains {not necessary 
cycles) obeying some transversality conditions. Then 
$\rho _a$ and $\rho _b$ are in general 
preobservables. This fact follows immediately from the 
considerations of \cite {[3]}.

Notice that the action of the group $Diff (S^1)$ of 
	orientation preserving diffeomorphims of
circle $S^1$  determines an action of this group on 
$LM$.  Factorizing $LM$ with respect to
this action we obtain a space $SM_{new}$that is 
homotopically 
equivalent to $SM$ .(This follows from
the fact that  $Diff (S^1)$ is homotopically equivalent 
to $S^1$.)   Similarly, instead of $LM$ we can 
consider a space $LM_{new}$ obtained from $LM$
by means of factorization with respect to the 
contractible group 
$Diff_0 S^1$ defined as a  subgroup of
$Diff (S^1)$  consisting of maps leaving intact the 
point $1\in \partial D$. 

{\bf  4. $A$-model and string bracket.} 

In this section we review some results of Fukaya \cite 
{[5]}. We 
will give also modification of 
these results to the form that allows us to relate them 
with the constructions of \cite {[3]}.

Let us consider a symplectic manifold $V$ and a 
Lagrangian submanifold $M\subset V$. 
Correlation functions of $A$-model on $V$ can be 
calculated by means of localization to 
moduli spaces of (psedo)holomorphic maps of Riemann 
surfaces;  in the case of open 
strings one should  consider maps of bordered surfaces 
transforming the boundary into 
$M$ \cite {[16]}. We restrict ourselves to  the genus 
zero case; 
then one should consider 
holomorphic maps $\varphi$  of the disk $D$ into $V$ 
obeying $\varphi (\partial D)\subset M$.
Every such map specifies an element of $\pi _2(V,M)$. 
One denotes by $\tilde {{\cal M}}(M, \beta)$ 
the moduli space of  holomorphic maps $\varphi: 
(D,\partial D)\rightarrow(V,M)$ that have a 
homotopy type $\beta \in \pi_2(V,M)$. We use the 
notations  $\hat {\cal M}(M,\beta)=\tilde {\cal M}
(M,\beta)/{\rm Aut}(D^2,1)$  and ${\cal M}(M,\beta)/PSL 
(2,{\bf R})$  where $PSL(2,{\bf R})$ is 
the group of fractional linear transformations  
identified with biholomorphic maps $D\rightarrow D$ 
and ${\rm Aut}(D,1)$ denotes its subgroup consisting of 
maps leaving intact the point $1\in \partial D$. 
The spaces   $\hat {\cal M}(M,\beta)$ and   $ {\cal 
M}(M,\beta)$ should be compactified by including 
stable maps from open Riemann surfaces of genus $0$; we 
will use the same notation for compactified 
spaces. 

Notice that   $ {\cal M}(M,\beta)$ specifies a chain 
${\cal M}_{\beta}$ in the string space $SM$. (We 
define a map  ${\cal M}(M,\beta)\rightarrow SM_{new}$ 
restricting every map $\varphi : D\rightarrow V$ 
belonging to   ${\cal M}(M,\beta)$ to the boundary of 
the disk $D$. We use in this construction 
the modified definition of $SM$ discussed at the end of 
Sec. 3. To obtain a chain in $SM$ we use a map of 
$SM_{new}$ onto $SM$
that specifies homotopy equivalence of these two spaces. 
)
 Similarly,  $\hat {\cal M}(M,\beta)$ 
specifies a chain $\hat {{\cal M}}_{\beta}$  in the loop 
space $LM$; the chain $\hat {{\cal M}}_{\beta}$
can be considered as a lift of ${\cal M}_{\beta}$. 
(Again we are using modified definition of $LM$ at the 
intermediate step.)
  
Fukaya \cite {[5]}, \cite {[6]} proved the following 
relation

\begin {equation}
\partial \hat {M}_{\beta}+{1\over 2}\sum_{\beta =\beta 
_1+\beta_2} \{ \hat{M}_{\beta_1},\hat {M}_{\beta _2} 
\}=0
\end {equation}
where $\{ , \}$ stands for the loop bracket in $LM$. We 
will derive from (8) the relation
\begin {equation}
\partial  M_{\beta}+{1\over 2}\sum_{\beta =\beta 
_1+\beta_2}[ M_{\beta_1}, M_{\beta _2} ]=0
\end {equation}
where $[ , ]$ denotes the string bracket in $SM$.

The derivation is based on relation  $\hat {{\cal 
M}}_{\beta} ={\rm lift}{\cal M}_{\beta}$. We notice that 
\begin {equation}
\partial \hat {M}_{\beta}=\partial ({\rm lift} 
M_{\beta})={\rm lift} (\partial M_{\beta})
\end {equation}
From the other side
$$\partial \hat {{\cal M}}_{\beta}=-{1\over 2}\sum_ 
{\beta_1+\beta_2=\beta} \{ \hat {{\cal M}}_{\beta_1}, 
\hat {{\cal M}}_{\beta_2}\}$$
$$=-{1\over 2} \sum_{\beta_1+\beta_2=\beta}\{ {\rm 
lift}{\cal M}_{\beta_1}, {\rm lift}{\cal 
M}_{\beta_2}\}$$
$$=-{1\over 2} \sum _{\beta_1+\beta_2=\beta} 
((-1)^{({\rm 
dim}{\cal M}_{\beta_1 +1)}} \Delta ({\rm lift}
{\cal M}_{\beta_1}\bullet {\rm lift}{\cal 
M}_{\beta_2})-(-1) ^{({\rm dim}{\cal M}_{\beta_1}+1)} 
\Delta({\rm lift}{\cal M}_{\beta_1}) 
\bullet {\rm lift}{\cal M}_{\beta_2}$$
$$-({\rm lift}{\cal 
M}_{\beta_1})\bullet \Delta ({\rm lift}{\cal 
M}_{\beta_2})$$ 
\begin {equation}
={\rm lift} (-{1\over 2}\sum _{\beta_1+\beta_2=\beta} 
[{\cal M}_{\beta_1}, {\cal M}_{\beta_2}])
\end {equation}

In the derivation of this formula we used (5), (6), (7) 
and 
relations $\Delta\bullet{\rm lift}=0, \ \ \Delta= {\rm 
lift} \bullet {\rm proj}$.
   
We obtain (9) comparing (8) and (11).
 
Let us fix a ring $\Lambda$ and a map $\alpha: H_2(V,M) 
\rightarrow \Lambda$ obeying $\alpha(\beta_1+\beta_2)
=\alpha(\beta _1)\cdot \alpha (\beta_2).$

We can construct a $\Lambda$-valued chain ${\cal M}$ on 
$SM$ taking
\begin {equation}
        {{\cal M}}=\sum _{\beta} \alpha_{\beta}  
{{\cal M}}_{\beta}. 
\end {equation}

It follows immediately from (11) that 
\begin {equation}
\partial {\cal M}+{1 \over 2}[{\cal M}, {\cal M}]=0.
\end {equation}

Usually one takes as $\Lambda$ the Novikov ring (a ring 
of formal expressions of the form $\sum a_i
T^{\lambda_i}$  where $a_i\in {\bf R}, \ \  \lambda_i\in 
{\bf R}, \lambda_i  \rightarrow +\infty$.) The 
map $\alpha$ should be fixed in a  way that guarantees 
finiteness of all relevant expressions. Our 
considerations will be completely formal; we refer to 
\cite {[7]} for an appropriate choice of $\alpha$.

{\bf 5. $A$-model and  Chern-Simons theory}

Let us start with the  chain ${\cal M}$ on $SM$ 
constructed at the end of Sec. 4.

We can construct the corresponding preobservable of 
generalized Chern-Simons theory using (13). 
It  follows immediately from (7) and (13)  that the 
preobservable $\rho =\rho_{\cal M}$ obeys 
$$\delta \rho +{1\over 2} \{ \rho ,\rho \} =0$$
We can modify the   Chern-Simons functional adding 
$\rho$. The new functional ${\cal S}+\rho$
verifies
$$\{ {\cal S} +\rho,{\cal S}+\rho \}=0. $$
This means that ${\cal S}+\rho $ can be considered as a 
solution of classical master equation (an 
action functional in BV formalism ) if ${\rm dim} M$ is 
odd and as a BRST generator 
if ${\rm dim} M$ is even. In the case ${\rm dim} M=3$  
the functional $\rho$ represents instanton 
corrections to the  Chern-Simons action;one can argue 
that this is true in any dimension.

 The 
above consideration is not  completely rigorous. We used 
the results of \cite {[2]}, \cite {[3]}, \cite {[4]} 
about the 
 string bracket on the space of chains in $SM$. These  
papers use different definitions of  string 
bracket; all of them agree on homology, however, it is 
essential for us to consider the bracket 
of chains that are not necessarily cycles. To give a 
rigorous proof one has to check that all 
results we are using can be verified with the same 
definition of  string bracket; this should not 
be a problem. 
 
We have seen that $A$-model instanton  corrections  to  
Chern-Simons functional  can be 
generalized very naturally to any dimension. This is a 
strong indication  that 
Chern-Simons functional  by itself also appears in 
multidimensional $A$-model. Indeed, analyzing Witten's 
arguments \cite{[16]} based on the application of string 
field theory one can reach a conclusion that $A$-model 
on $T^*M$ is equivalent to the generalized 
Chern-Simons theory on $M$.
(One can understand from Witten's paper, that he was 
aware of possibility of multidimensional generalization 
of his constructions.)

 It seems that the mathematical 
justification of this statement can be based on the idea 
that a stack of $N$ coinciding D-branes can be replaced 
by $N$ Lagrangian submanifolds that depend on some 
parameter and coincide when the parameter tends to $0$. 
This situation was studied by Fukaya-Oh \cite {[8]} and 
Kontsevich-Soibelman \cite {[11]}. 

Let us consider $N$ transversal Lagrangian submanifolds 
$M_1,.,M_N$ in symplectic manifold $V$. One can 
construct 
corresponding $A_{\infty}$-category (Fukaya category) 
\cite {[7]}. 
The construction of operations in this category is 
based on the consideration of moduli spaces of 
pseudoholomorphic maps of a disk $D$ into $V$. (One  
assumes that $V$ is equipped with almost complex 
structure $J$; in the case when $V=T^*M$ one assumes 
that almost complex structure is induced by a   metric 
on $M$.) One fixes the intersection points $x_i\in 
M_i\cap 
M_{i+1}$ for $1\leq i \leq N-1$ and $x_N\in M_N \cap 
M_1$. 
The Fukaya category is defined in terms of moduli spaces 
$M_J^z 
(V,M_i,x_i)$ of $J$-holomorphic maps $v: D\rightarrow V$ 
transforming given points $z_i \in \partial D$ into 
points $x_i$.

One should consider also the union of all spaces 
${\cal M}_J^z$ where $z_i$ run over all cyclically 
ordered subsets of $\partial D$ and factorize this 
union with respect to the group $PSL(2,{\bf R})$ 
acting as a group of biholomorphic automorphisms 
of the disk; one obtains the moduli spaces 
${\cal M}_J(V,M_i,x_i)$.
The definition of operations in Fukaya category involves 
summation over ${\cal M}_J$. 

Following \cite {[8]} we can consider 
the case when $V=T^*M$ and the Lagrangian submanifolds 
$M_i$ are defined as graphs $M_i ={(x,\xi)\in 
T^*M|\xi=\varepsilon df_i(x)}$ where $f_1,.,f_N$ are 
such functions on $M$ that difference between any two of 
them 
is a Morse function; then the corresponding Lagrangian 
submanifolds are transversal and intersection points 
$x_i\in M_i\cap M_{i+1}$ are critical points of 
functions 
$f_i-f_{i+1}$. Fukaya and Oh \cite {[8]}
have studied the moduli spaces ${\cal M}_J(V, M_i,x_i)$
for this choice of Lagrangian submanifolds. They have 
proved 
that for small $\varepsilon$ these  moduli spaces are 
diffeomorphic to  moduli spaces ${\cal M}_g(M,f_i,p_i)$
of graph flows. (An element of moduli spaces 
${\cal M}_g(M,f_i,p_i)$ where $p_i$ are critical points 
of 
$f_i-f_{i+1}$ is a map of a metric graph $\gamma$ into 
$M$ 
transforming edges of the graph $\gamma$ into  
trajectories of negative gradient flow of the difference 
of 
two of the functions. It is assumed that the graph 
$\gamma$ 
is a rooted tree embedded into the disk $D$ and the 
exterior 
vertices are mapped into $\partial D$.)

This picture is very close to the Witten's picture \cite 
{[16]}
where graphs appear as  degenerate instantons. It is 
clear from it that $A$-model on $T^*M$ can be reduced  
to quantum field theory-summation over 
embedded holomorphic  disks can be replaced by the 
summation over graphs. However, it is not clear yet 
that this quantum field theory coincides with 
Chern-Simons theory. To establish this one can apply 
the results of \cite {[11]}. 

The papers \cite {[8]} and \cite {[11]} use the 
language of $A_{\infty}$-categories. In this 
language the results of \cite {[8]} can be formulated in 
the 
following way: Fukaya $A_{\infty}$-category constructed 
by 
means of Lagrangian submanifolds of $T^*M$ is equivalent 
to 
Morse $A_{\infty}$-category of smooth functions on $M$. 
It is 
proved in \cite {[11]} under certain conditions that the 
Morse 
$A_{\infty}$-category is equivalent to de Rham category. 
All 
$A_{\infty}$-categories (or, more precisely, 
$A_{\infty}$-precategories) in question are equipped 
with 
inner product; the equivalence is compatible with inner 
product.

The minimal model of Fukaya $A_{\infty}$-category is 
related 
to tree level string amplitudes; the relation of these 
amplitudes 
to Chern-Simons theory can be derived from the remark 
that quasiisomorphic $A_{\infty}$-algebras with inner 
product specify equivalent Chern-Simons theories. 

It is important to emphasize that $A$-model for any 
genus 
is related to Chern-Simons theory. It was 
mentioned 
in \cite {[8]} that not only moduli spaces of 
pseudoholomorphic 
disks 
on $T^*M$ but also moduli spaces of higher genus 
pseudoholomorphic curves can be described in terms of 
graphs. Again, this is consistent with equivalence of 
$A$-model  to quantum field theory. In 
simplest case the relation to Chern-Simons theory was 
studied 
in \cite {F}.

{\bf Acknowledgments.} I am indebted to K. Fukaya,
M. Kontsevich, M.Movshev, Y.G. Oh and Y. Soibelman  
for interesting discussions. I appreciate the
hospitality of KIAS where I learned from Y.G. Oh
about unpublished results by K. Fukaya used in this
paper.

\end {document}